\documentclass[aps,prl,reprint,superscriptaddress]{revtex4-2}
\usepackage{amsmath}
\usepackage{amssymb}
\usepackage{bm}   % bold math
\usepackage{graphicx}   % Include figure files
\usepackage{upgreek}   % for regular (not italic) greek symbols
\usepackage[colorlinks,allcolors=black]{hyperref}

\begin{document}

%\preprint{}

\title{Efficient Navigation of Colloidal Robots in an Unknown Environment via Deep Reinforcement Learning}


\author{Yuguang Yang}
%\thanks{}
\affiliation{Institute of Biomechanics and Medical Engineering, Applied Mechanics Laboratory, Department of Engineering Mechanics, Tsinghua University, Beijing 100084, China}
\affiliation{Chemical and Biomolecular Engineering, Johns Hopkins University, Baltimore, MD 21218}

\author{Michael A. Bevan}
\affiliation{Chemical and Biomolecular Engineering, Johns Hopkins University, Baltimore, MD 21218}

\author{Bo Li}
\email[]{libome@tsinghua.edu.cn}
% \thanks{}
\affiliation{Institute of Biomechanics and Medical Engineering, Applied Mechanics Laboratory, Department of Engineering Mechanics, Tsinghua University, Beijing 100084, China}


\date{\today}

\begin{abstract}
Equipping active colloidal robots with intelligence such that they can efficiently navigate in unknown complex environments could dramatically impact their use in emerging applications like precision surgery and targeted drug delivery. Here we develop a model-free deep reinforcement learning that can train colloidal robots to learn effective navigation strategies in unknown environments with random obstacles. We show that trained robot agents learn to make navigation decisions regarding both obstacle avoidance and travel time minimization, based solely on local sensory inputs without prior knowledge of the global environment. Such agents with biologically inspired mechanisms can acquire competitive navigation capabilities in large-scale, complex environments containing obstacles of diverse shapes, sizes, and configurations. This study illustrates the potential of artificial intelligence in engineering active colloidal systems for future applications and constructing complex active systems with visual and learning capability.
\end{abstract}

%\keywords{}

\maketitle

% Introduction
Self-propelled active colloidal particles have recently demonstrated great promise as micro-robots capable of functioning in complex confined and crowded micro-environments. In potential real-world applications (e.g., drug delivery, precision surgery, search, and environmental remediation \cite{JLi_SciRobot_2017,Mallouk_SciAm_2009,JLi_ACSNano_2016,Soler_ACSNano_2013,Volpe_PNAS_2017}), micro-robots are confronted with navigation challenges including long-distance travel (e.g., tissue, soil, and vasculature), unknown or spatiotemporally changing environment abundant with obstacles and dead-ends, and additional time and fuel constraints. Beyond developing sophisticated micro-robot systems that have more efficient transport mechanisms and  sensing capabilities  \cite{Mallouk_SciAm_2009,Sanch_AngChem_2015,Koman_NatNano_2018, Wu_SciRobot_2019}, efforts have also been directed towards developing better navigation strategies \cite{Haeufle_PRE_2016,Liebchen_arXiv_2019,Qian_Chem_2013,Yang_ACSNano_2018}. Examples include application of a Markov decision process framework \cite{Yang_ACSNano_2018} and a variational Fermat's principle \cite{Liebchen_arXiv_2019} to compute optimal navigation paths in mazes and flow fields. However, these methods generally require pre-existing knowledge of the entire environment, which is generally unavailable in the vast majority of expected applications, and even when available, the computational cost becomes prohibitive for large scale navigation.

\begin{figure}[h]
\centering
\includegraphics[scale=0.6]{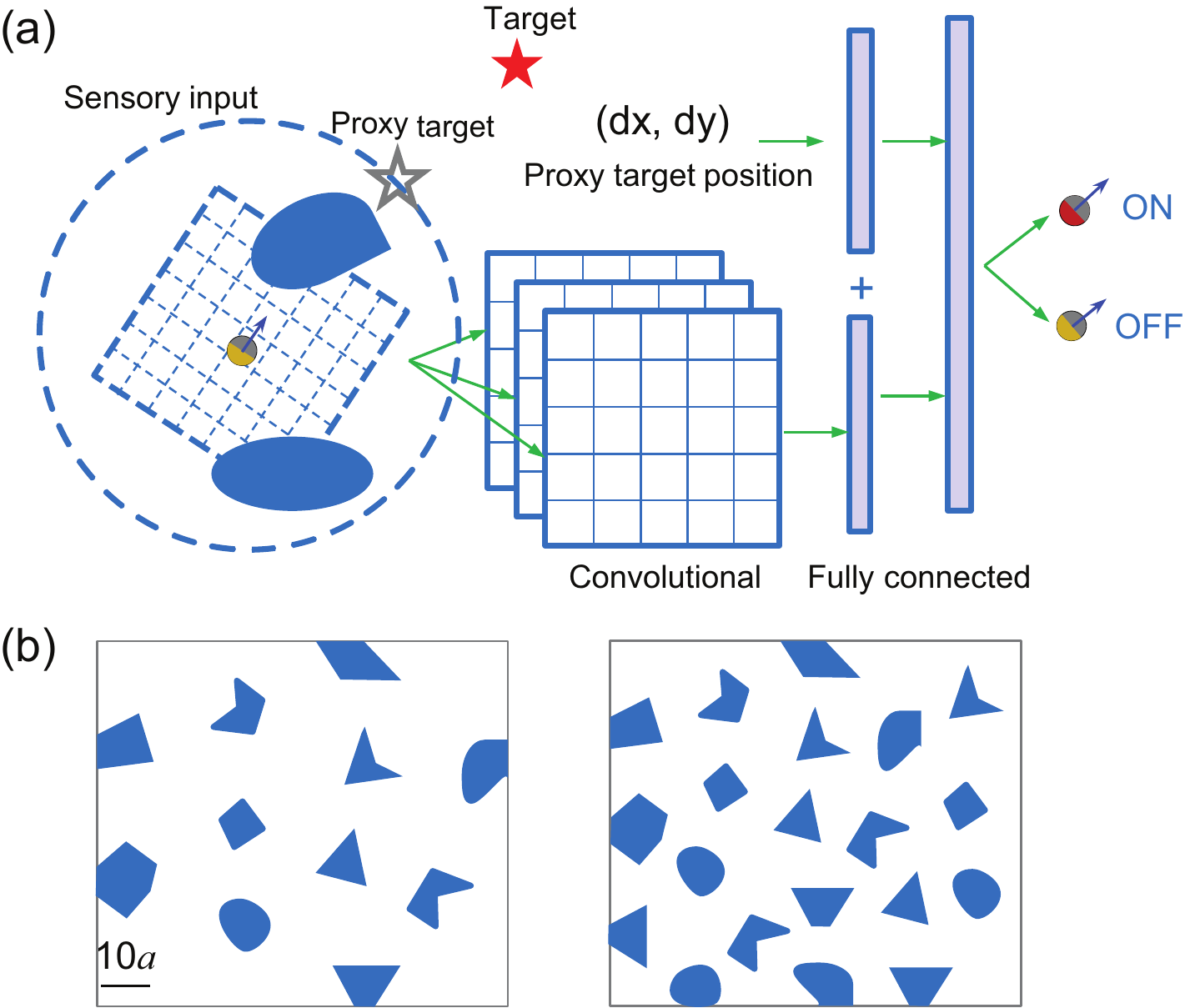}
\caption{\label{Fig_1}(a) The neural network architecture used in our DRL algorithm. The details of the architecture are given in the Supplemental Material \cite{SM}. Two streams of sensory inputs are fed to the neural network, including pixel image ($30a \times 30a$) of the particle’s neighborhood fed into convolutional layers and the target’s position fed into a fully connected layer. A distant target will be transformed to a local proxy target. The network will output two values corresponding to the Q values of the ON and OFF actions. (b) The two obstacle environments used to train the particle agent.}
\end{figure}

Animals are capable of effortlessly navigating and exploring in unknown, complex, visually rich environments (e.g, cities, fields, mountains, sea, space, etc.) \cite{Aidley_1981}. An adult in an unfamiliar city has little difficulty traversing city blocks and exploiting short-cuts when possible. In the context of colloidal robot navigation in an unknown obstacle field, here we aim to overcome navigation challenges by equipping colloidal robots with human-like decision capabilities, or artificial intelligence, which has achieved remarkable recent success in diverse problems \cite{Komo_NatMed_2018,Levine_JMLR_2016,Lillic_arXiv_2015,Mnih_Nature_2015,Popova_SciAdv_2018,Silver_Nature_2016,Verma_PNAS_2018,Zhou_ACSCentSci_2017}. We use a neural network architecture \cite{LeCun_Nature_2015} that aims to minimally mimic the human navigation decision making system in two key aspects [Fig. \ref{Fig_1}(a)]. First, we use convolutional neural layers \cite{LeCun_Nature_2015,Kriz_2012}, which mimics the human visual system \cite{Serre_IEEE_2005} by directly processing high-dimensional raw sensory information (i.e. obstacle images). Raw sensory information has been showed to enable learning of efficient representations from high-dimensional observations and facilitates the generalization of knowledge to new situations (i.e., unknown environments) \cite{LeCun_Nature_2015,Mnih_Nature_2015}. Second, we dynamically transform distant targets to local short-range proxy targets, which mimics human navigation behaviors that decompose long-range goals into a series of shorter-range sub-goals (on the scale of visual perception limit) \cite{McComb_JEB_2008,Mous_PNAS_2011,Barb_PRL_2016}.  

We train the neural network [Fig. \ref{Fig_1}(a)] via a model-free learning framework, which is based on algorithms collectively referred to as deep reinforcement learning (DRL). We deploy the robot agent to navigate diverse environment structures (e.g., obstacle shapes, sizes, spacing, and target distances) [Fig. \ref{Fig_1}(b)] and train the neural network directly from extensive navigation trajectory data, involving sequences of observations, self-propulsion decisions, subsequent observations, and reward signals. Ultimately, the trained agent learns to make self-propulsion decisions to circumvent obstacles and reach targets, accommodating the unique stochastic dynamics of active particles. Because our DRL algorithm only relies on local information, the agent is inherently trained to acquire short-range navigation capability; however, the architecture [Fig. \ref{Fig_1}(a)] enables it to navigate long distance by implicitly decomposing them into a series of short-scale navigations.

In a series of navigation tests, we show in the following that the trained agent can efficiently navigate in large, complex environments with obstacles of unknown shape and arrangement. We demonstrate that the neural network can learn effective representations of observations and thus shed light on the successful navigation performance. Our results demonstrate a general framework to train an intelligent colloidal robot to master the rule of efficient navigation based on local visual information, in contrast to developing algorithms to compute navigation strategies on a case by case basis. Ultimately, such a DRL approach can be employed in future applications of colloidal robots or as a building block to construct complex active systems with visual and learning capability\cite{Charlesworth_PNAS_2019}.

We model colloidal robots as Brownian-type active particles that can self-propel via either internal or external energy injection yet without control over its orientation. The motion of such self-propelled particle confined on a plane is given by 
\begin{align}
\partial_t x & =\xi_x(t) + v\cos\theta, \notag \\
\partial_t y & =\xi_y(t) + v\sin\theta ,\\
\partial_t \theta&=\xi_{\theta}(t), \notag
\end{align}
where $x$, $y$, and $\theta$ denote the position and orientation, $t$ is time, and $v$ is propulsion speed taking binary values of 0 and $v_{\rm max}$ as the control inputs (denoted by OFF and ON). Brownian translational and rotational displacement processes $\xi_x$, $\xi_y$, and $\xi_{\theta}$ are zero-mean Gaussian noise process with variances $\langle \xi_x (t) \xi_x(t')\rangle =2D_{\rm t} (t-t')$, $\langle \xi_y(t) \xi_y(t')\rangle =2D_{\rm t}(t-t')$, and $\langle \xi_{\theta}(t) \xi_{\theta}(t')\rangle=2D_{\rm r}(t-t')$, respectively, where $D_{\rm t}$ is the translational diffusivity and $D_{\rm r}$ is the rotational diffusivity. All lengths are normalized by particle radius $a$ and time is normalized by $\tau= 1 / D_{\rm r}$. The control update time is $t_{\rm c} = 0.1\tau$, the integration time step $\Delta t =0.001\tau$, and $v_{\rm max} = 2a/ t_{\rm c}$. Note that orientation $\theta$ is subject to Brownian rotation and is uncontrolled.

We denote the particle state by $s_n = (x_n, y_n, \theta_n)$. The observation $\phi(s_n)$ at $s_n$ consists of the pixelated image of the particle's neighborhood and the target position $(x^t, y^t)$, as shown in Fig. \ref{Fig_1}(a).  We seek an optimal control policy, $\pi^*$, which maps an observation $\phi(s_n)$ to a self-propulsion action (i.e. ON or OFF) to maximize the expected reward accumulated during a navigation process, $\mathbb{E}\sum_{n=0}^{\infty}\gamma^n[R(s_n)]$, where $R$ is the one-step reward function and $\gamma$ is set to 0.99 to encourage the agent to value rewards in distant future. To seek an optimal policy minimizing arrival time \cite{Yang_ACSNano_2018,Sutton_1998}, $R$ is set equal to 1 for all states that are within a threshold distance 2 to the target and 0 otherwise. The optimal control policy is obtained by training the neural network [Fig. \ref{Fig_1}(a)] to approximate the optimal action-value function (known as the $Q$ function) given by
\begin{align}
Q^*(\phi(s),v)=&\mathbb{E}[R(s_0)+\gamma^1 R(s_1)+\gamma^2 R(s_2)+\cdots \notag \\
&|\phi(s_0)=\phi(s),v_0=v,\pi^*],
\end{align}
which is the expected sum of rewards along the process by following the optimal policy $\pi^*$, after observing $\phi(s)$ and self-propelling with speed $v$. The neural network contains convolution neural layers to process sensory information of the particle neighborhood, represented by a $W\times W$ binary image ($W = 30$), and a fully connected layer to process the target's position in the particle's local coordinate system. Distant targets (distance $> W$) are transformed into local proxy targets projecting onto a circle with radius $W$ [Fig. \ref{Fig_1}(a)] (see \cite{SM}). The neural network finally outputs the two $Q^*$ values associated with the ON and OFF actions.  

We use the canonical deep $Q$ learning algorithm \cite{Mnih_Nature_2015} to iteratively improve the the estimate of $Q^*$, with several enhancements \cite{Hasselt_2016,Andry_2017} to improve sample efficiency and the rate of convergence \cite{SM}. During training, extensive navigation data in free spaces and two obstacle-present environments [Fig. \ref{Fig_1}(b)] are collected to enable the agent to learn navigation strategies in various scenarios (different obstacle shapes, sizes, spacing, and target locations).  With the estimated $Q^*$ function, the optimal propulsion decision at a given observation $\phi(s)$ is given by $\pi^*={\rm argmax}_v Q^*(\phi(s), v)$.  

\begin{figure*}[ht]
\centering
\includegraphics[scale=0.65]{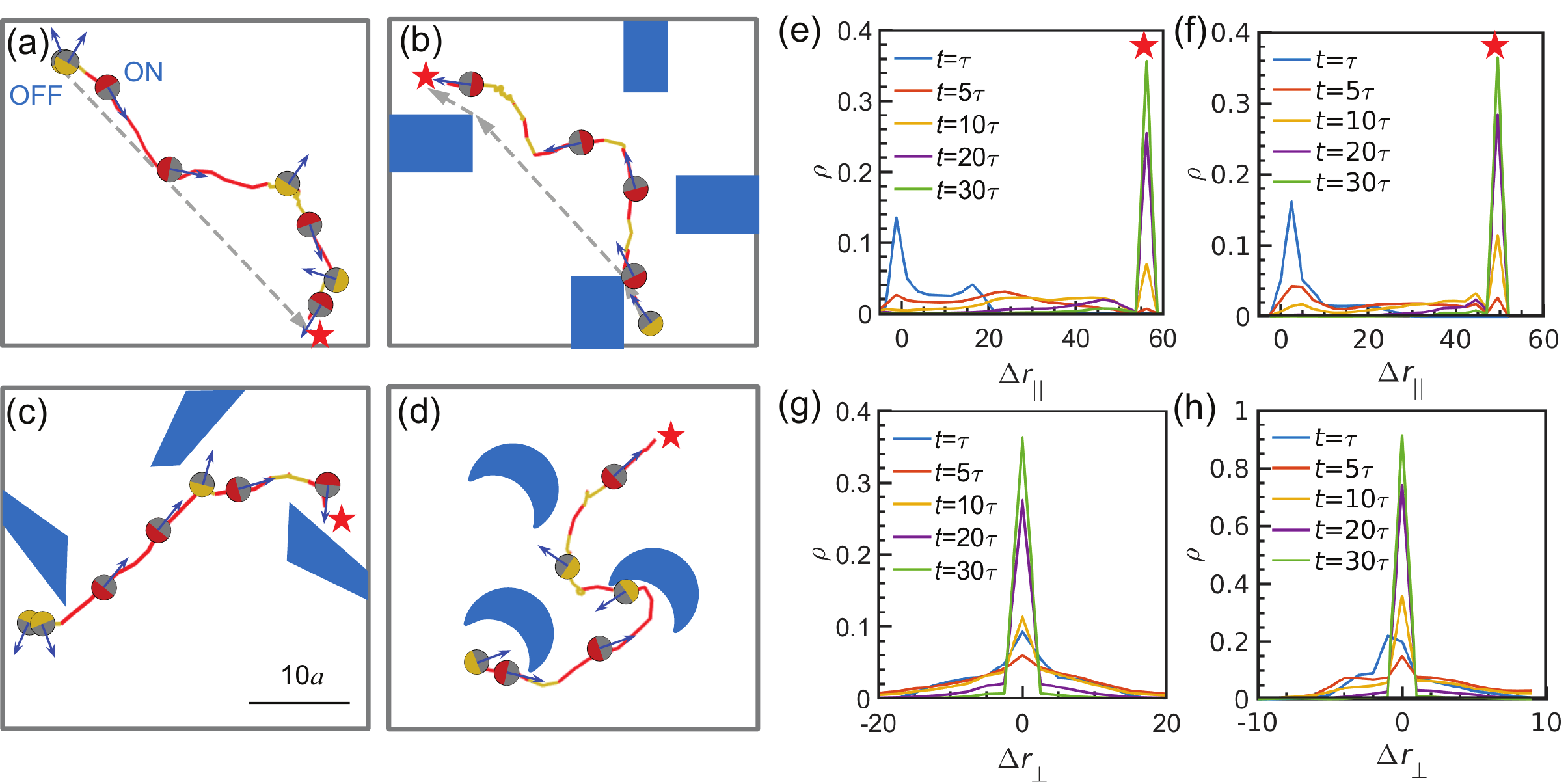}
\caption{\label{Fig_2}Navigation trajectories of DRL trained agent in different environments, including (a) free space, (b) rectangle obstacles, (c) reentrant rectangle obstacles, and (d) moon-shape obstacles. Targets are denoted by red stars. Trajectories are colored red when self-propulsion is turned ON and are colored gold when self-propulsion is turned OFF. Grey dash vectors are global shortest path from the starting point to the target. (e--h) The distributions of parallel and perpendicular components of displacements w.r.t. grey dash vectors when navigating in free space (e, g) and in rectangle-obstacle environment (f, h).}
\end{figure*}

The trained DRL active particle agent can efficiently navigate in free space and unknown obstacle-present environments (i.e. not observed before during its training process) [Fig. \ref{Fig_2}(a--d)].  In free space navigation [Fig. \ref{Fig_2}(a)], the navigation strategy derived from the learned $Q^*$ function is similar to previous studies \cite{Haeufle_PRE_2016,Qian_Chem_2013,Yang_ACSNano_2018} and can be summarized approximately as,
\begin{equation}
\pi^*(s)=\left\{ \begin{array}{rl}
v_{\rm max}, & d_n \in [d_{\rm c}, \infty), \\
v_{\rm max}, & d_n \in [0,d_{\rm c}),  \alpha_n \in [-\alpha_{\rm c},\alpha_{\rm c}], \\
0, & {\rm otherwise},
\end{array} \right.
\end{equation}
where $d_n$ is the projection of the target-particle vector onto the orientation vector $\mathbf{n}=(\cos \theta, \sin\theta)$, $\alpha_n$ is the angle between target-particle distance vector and $\mathbf{n}$, and parameter $d_{\rm c}$ and $\alpha_{\rm c}$ are fitted to be $\sim 0.4 v_{\rm max}\tau$ and $\sim 30^{\circ}$. Intuitively, the particle agent has learned an `orientation timing' strategy where it will propel itself when the target is located in front of the particle, or otherwise waits for the favorable orientation to be sampled by Brownian rotation.

We further consider navigating in environments with obstacles of different shapes and arrangements unseen in the training, including regular rectangle obstacles [Fig. \ref{Fig_2}(b)], reentrant rectangle obstacles [Fig. \ref{Fig_2}(c))], and moon-shaped obstacles [Fig. \ref{Fig_2}(d)]. These obstacles, with different shapes, size and spacing, are designed to block paths to targets, or trap particles via concave geometries (obstacle-wall arrangements [Fig. \ref{Fig_2}(b) and (c)], or obstacles geometry itself [Fig. \ref{Fig_2}(d)]). Successful navigation trajectories indicate the trained particle agent has learned navigation `knowledge' generalizing beyond the training environments (i.e., Fig. \ref{Fig_1}(b)). Specifically, the agent learns to wait for desired orientations to propel itself around general convex blocking obstacles and avoid dead-ends [Figs. \ref{Fig_2}(b) and (c)] or temporally propel itself away from concave traps [Fig. \ref{Fig_2}(d)].  In general, the agent learns to avoid the obstacles and approximately follow the shortest geometric path towards the target, even only local information is used.

We examine the distribution of displacements of the trained agent during the navigation processes in [Fig. \ref{Fig_2}(a) and (b)]. In collecting the statistics, trajectories start at the same initial position but with randomly sampled initial orientations. The displacements are projected along and perpendicular to the shortest geometric path (represented by the direction vectors connecting from the initial position to the target), which are used to capture the navigation progress and the deviation from the ideal direct path, respectively. As shown in Fig. \ref{Fig_2}(e) and (f), for navigation in both free space and rectangle-obstacle environment, the displacement distributions exhibit two modes at $t=\tau$: (i) a near mode located at 0 due to trajectories with unfavorable initial orientations are waiting at the initial position for favorable orientations from stochastic rotation; (ii) a far mode located at $\sim 0.75 v_{\rm max}\tau$ due to trajectories with favorable initial orientation rapidly self-propelling. At longer times, the near mode continues to spread out and the far mode propagates towards the target, indicating that the trained agent can readily propel itself to get closer to the target when favorable orientations and positions appear via Brownian motion. In addition, the perpendicular displacement (i.e., the vertical deviation to the ideal path) distributions exhibit a narrow peak around 0 but with tails extending to $\sim v_{\rm max}\tau$ in free space and $\sim 0.5 v_{\rm max}\tau$ in obstacle-present environment, indicating the DRL trained agent can usually maintain a close distance to the ideal path but occasional large deviation can also occur [Fig. \ref{Fig_2}(g) and (h)].  

\begin{figure}[t]
\centering
\includegraphics[scale=0.55]{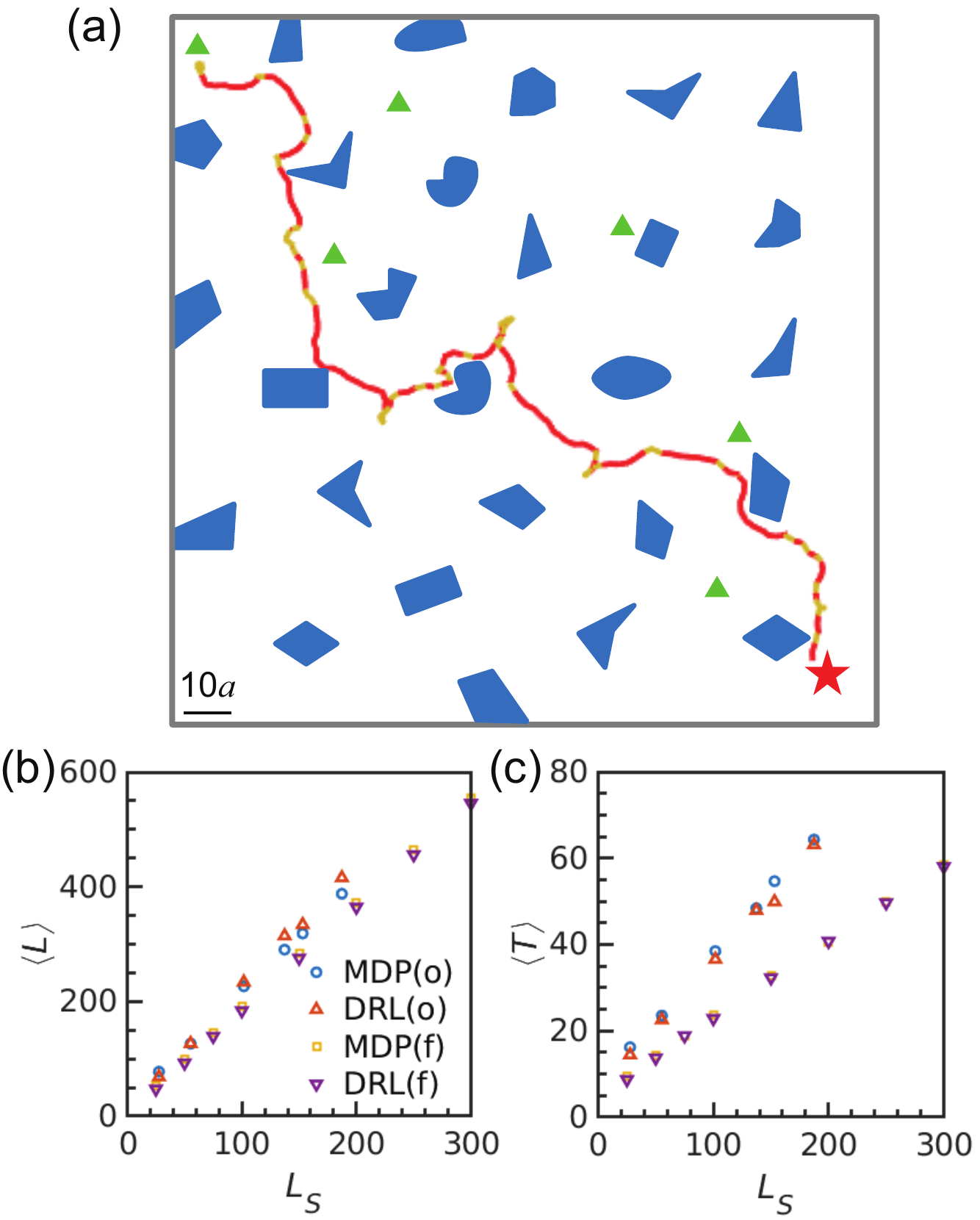}
\caption{\label{Fig_3}(a) An unknown large-sized test environment used for performance benchmark. Solid curve shows an example trajectory of a DRL trained agent navigating from the upper left starting point to the target. (b--c) Mean traveled path length $\langle L \rangle$ and mean first arrival time $\langle T \rangle$ are compared for a DRL and a MDP trained agent navigating in free space (denoted by DRL(f) and MDP(f)) and in obstacle environment(denoted by DRL(o) and MDP(o)).  Navigations from different starting points (green symbols in (a)) to the prescribed target are analyzed. These different navigation tasks can be parameterized by the shortest geometric path length $L_s$ from starting points to the target.}
\end{figure}

We now benchmark the navigation performance across different length scales [Fig. \ref{Fig_3}(a)] with the model-based Markov decision process (MDP) algorithm \cite{Yang_ACSNano_2018,Puter_2014,SM}, which can compute the optimal navigation strategy using a discretized particle dynamic model. MDP cannot be applied to unknown environment navigation (as it requires pre-knowledge of the environment) or large-scale navigation tasks (prohibitive computational cost). The performance of DRL and MDP algorithms are compared based on mean traveled path length $\langle L \rangle$ and mean first arrival time $\langle T \rangle$. The traveled path length of a trajectory is defined by $L=\sum_{i=1}^{N-1}\sqrt{(x_{i+1}-x_i)^2+(y_{i+1}-y_i)^2}$  , where $(x_1, y_1)$, $(x_2, y_2)$, $\cdots$ $(x_N, y_N)$ are particle positions observed at consecutive control update times. For trained agents navigating different distances in both free space and the obstacle-present environment [Fig. \ref{Fig_3}(a)], DRL trained agent and MDP trained agent perform similarly in terms of traveled path length [Fig. \ref{Fig_3}(b)] and arrival time [Fig. \ref{Fig_3}(c)]. Since the MDP agent is known to seek global optimal performance by planning the global shortest paths according to the pre-existing map, the comparable performance indicates the DRL agent can learn competitive strategies based solely on local information. The DRL agent's average travel speeds $v=\langle L \rangle/\langle T \rangle$ are fitted to $\sim 0.36 v_{\rm max}$ for navigation in the obstacle environment, smaller than the fitted result of $\sim 0.5 v_{\rm max}$ in free space, since the existence of obstacles prevents particles from propelling as frequently as in free space. 

\begin{figure}[t]
\centering
\includegraphics[scale=0.55]{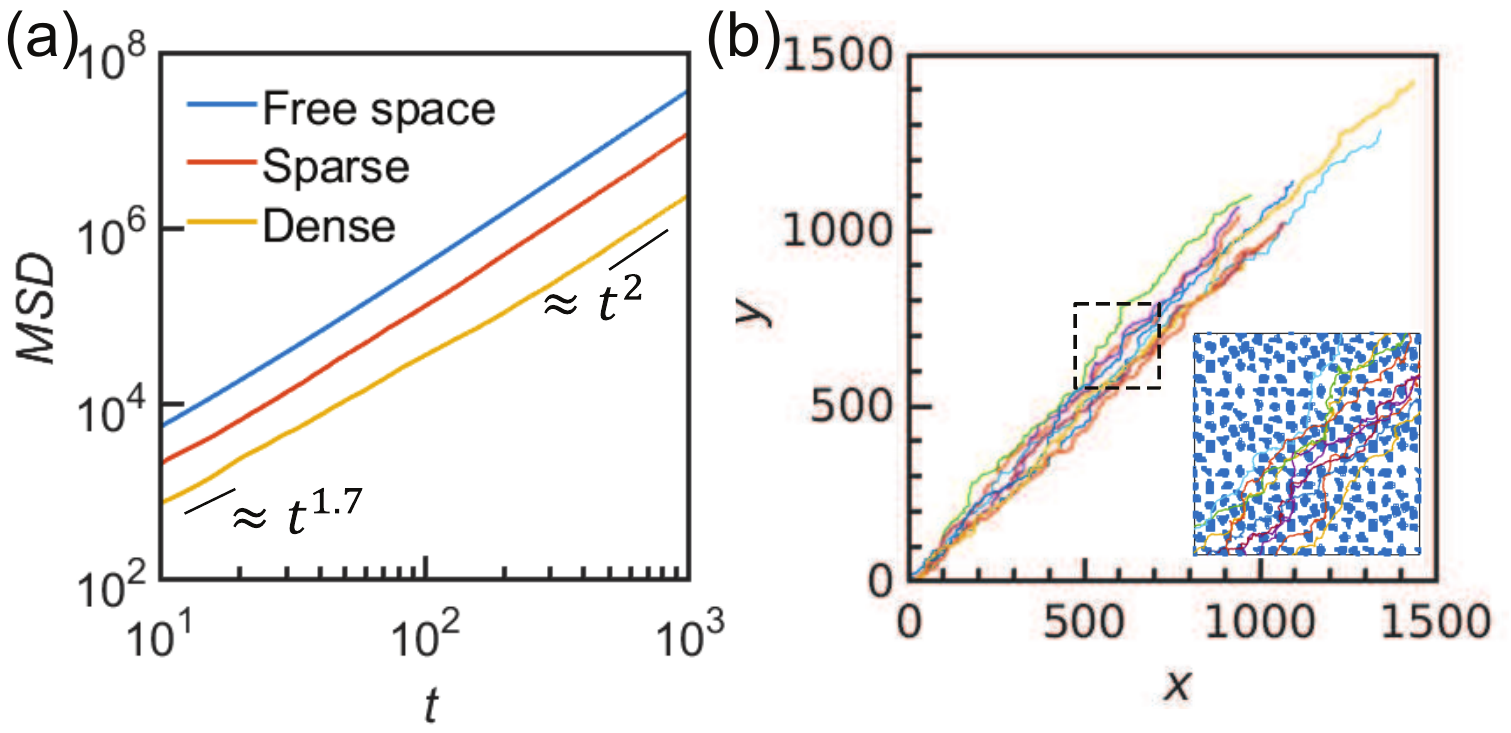}
\caption{\label{Fig_4}(a) The mean squared displacement (MSD) for a DRL trained agent in free space, and two unknown environments with sparse obstacles and dense obstacles. (b) Ten representative trajectories (lasting 1000$\tau$) during directed transport along $45^{\circ}$ direction in a periodic dense obstacle environment (obstacles not shown). Inset figure is the magnified dashed region. Also see Fig. S3 in \cite{SM}.}
\end{figure}

To ultimately evaluate the navigation capability in infinitely-large environments, we examine the mean squared displacement (MSD) for particles performing directed transport in free space, and the periodic environments contains sparse obstacles [Fig. \ref{Fig_4}(a)], and dense obstacles (see Fig. S3 in \cite{SM}). At both shorter and longer time scale $t$ compared to $\tau$, the MSD of the trained agent particle navigating in free space follows ${\rm MSD}(t) \sim t^2 $ [Fig. \ref{Fig_4}(a)], as the navigation strategy produces intermittent directed transport. Surprisingly, for navigations in the obstacle environment, the DRL trained agent still displays approximately ${\rm MSD}(t) \sim t^2 $ at longer time scale, although at smaller time scale it displays ${\rm MSD}(t) \sim t^{1.7}$, where the smaller exponent arises from detours when circumventing obstacles. For the same reason, the transition from $t^{1.7}$ occurs at longer time for environments with dense obstacles ($\sim 100\tau$  v.s. $\sim 10\tau$). The similar scaling of $t^2$ in both free space and obstacle environment can also be illustrated by the  representative trajectories of directed transport in an obstacle environment [Fig. \ref{Fig_4}(b)], where the DRL agent manages to circumvent all obstacles and maintains a close distance to the ideal directed path (i.e. ray in the $45^{\circ}$ direction).  In short, the trained agent can perform directed transport in obstacle-present environments as in obstacle-absent ones (i.e., without getting trapped), although at a reduced speed.

\begin{figure}[t]
\centering
\includegraphics[scale=0.48]{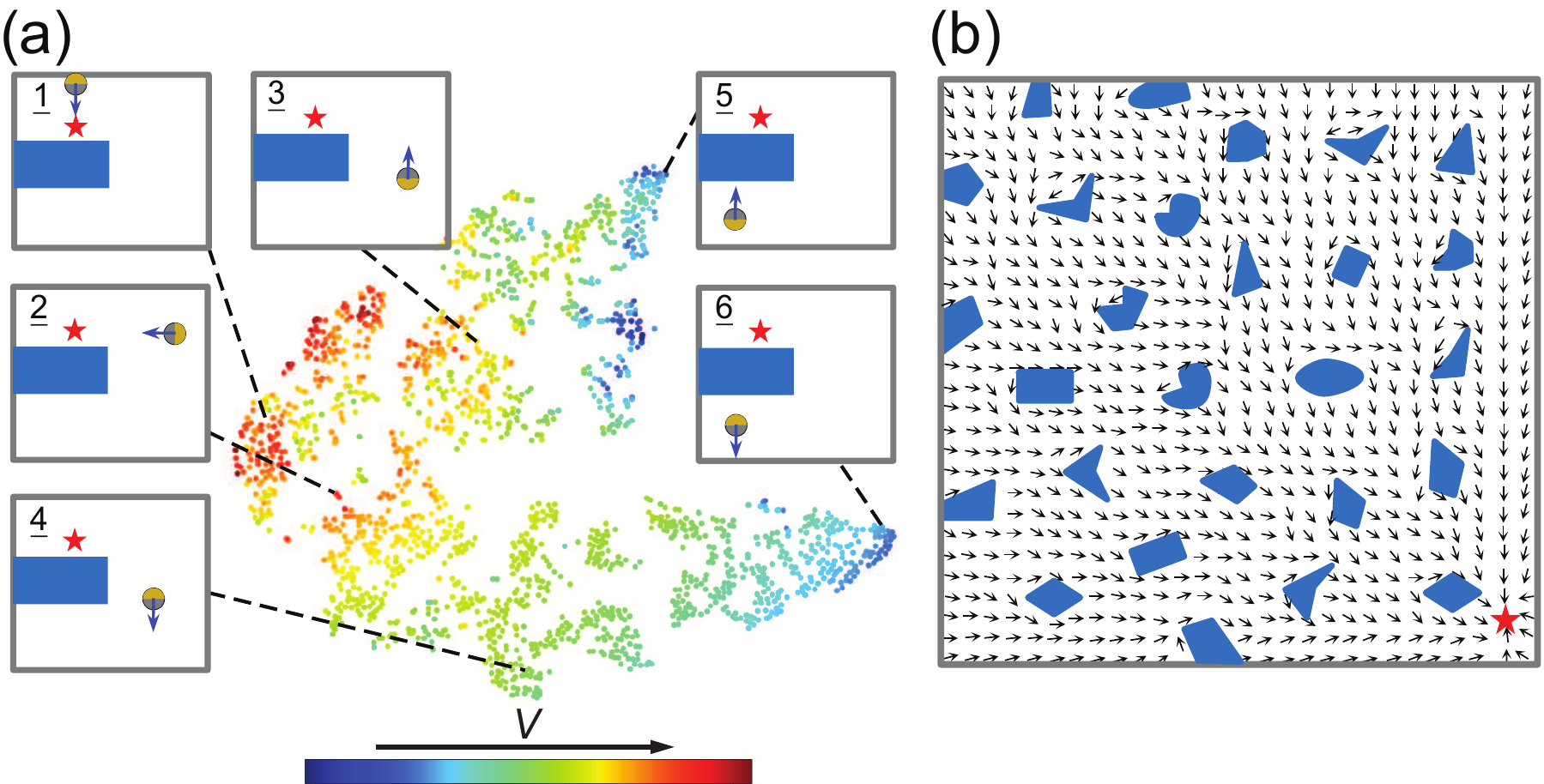}
\caption{\label{Fig_5}(a) Two-dimensional t-SNE embedding of the representation in the last hidden layer of the neutral network in a single obstacle navigation task. Every point corresponds to 2D representation of observations at admissible particle states $(x, y, \theta)$ discretized with step size $(1, 1, \pi/8)$. Points are colored by state value $V= {\rm max}_v Q^*(\phi(s), v)$ (Eq. \eqref{eqV}). (b) Implied effective navigation direction from the gradient of orientation averaging optimal state value (Eq. \eqref{eqVxy}) with the target denoted by red star.}
\end{figure}

Finally, we examine the representations learned from high dimensional sensory inputs by the neural network to understand the successful performance in the previous navigation tasks. We consider an unknown environment with only one obstacle and apply the t-Distributed Stochastic Neighbor Embedding (t-SNE) algorithm \cite{Maaten_JMLR_2008} to embed the learned representations in the last hidden layer into a 2D plane [Fig. \ref{Fig_5}(a)]. Each 2D point is colored by the state value defined by 
\begin{equation}
V(s)={\rm max}_v Q^*(\phi(s),v)=\mathbb{E}\left[\sum_{n=0}^{\infty}R(s_n)| s_0=s, \pi^* \right], \label{eqV}
\end{equation}
where $V(s)$ can be interpreted as the expected maximum total reward a particle can achieve when it initializes with state $s$ and follows the optimal navigation policy $\pi^*$. A particle state with higher $V$  indicates a particle starting at this state can arrive at the target faster than other states with lower $V$ .  

As shown in Fig. \ref{Fig_5}(a), high dimensional sensory observations at different particle states are embedded in the 2D plane apparently based on the shortest path distance to the target location (in the horizontal direction) and the particle’s orientation (in the vertical direction). In the left end, particle states with the shortest distance and favorable orientations are assigned with highest state value [Fig. \ref{Fig_5}(a$_1$)], while particle states with larger distance are assigned with lower state values [Fig. \ref{Fig_5}(a$_2$)]. Particle states with intermediate range to the target are placed in the middle with intermediate state values and locate at the upper part if the particle favorably orients to the target [Fig. \ref{Fig_5}(a$_3$)] or lower part if the colloidal particle orients opposite to it [Fig. \ref{Fig_5}(a$_4$)]. For particles located the furthest from the target due to the blockage of the obstacle, their observations are assigned to the lowest state value and placed on the rightmost end at either upper or lower wings [Fig. \ref{Fig_5}(a$_5$) and (a$_6$)]. While $Q$ function assigns a lower value to states with unfavorable orientations for particles within an intermediate range, it assigns similar state values to distant particle states regardless of their orientations; this is because initial orientation does not add appreciable value to long-distance navigation. In short, as demonstrated in the toy example [Fig. \ref{Fig_5}(a)], the reward signals have shaped the neural network to learn representations from observations that can distinguish whether one particle state is more favorable than the other via state values.

Another way to understand the learned navigation strategy is by visualizing the normalized gradient vector of orientation averaging optimal state values $V_{XY}(x,y)$ defined by \cite{SM}
\begin{equation}
V_{XY}(x,y)=\frac{1}{8}\sum_{i=1}^8 V(x,y,\frac{i\pi}{4}). \label{eqVxy}
\end{equation}

Because the trained agent will take actions to move from low value states (e.g., particle states with unfavorable orientation and far away from the target) to high value states (e.g., particles states with favorable orientation and smaller distance to the target), the gradient of $V_{XY}(x,y)$ represents the effective navigation direction that the trained agent tends to move. In Fig. \ref{Fig_5}(b), the effective navigation direction in the complex test environment clearly shows the agent’s intent to circumvent all obstacles (even they are not observed by the agent before) and follow paths directed towards the target, which is consistent with navigation trajectories observed in Fig. \ref{Fig_3}(a) and Fig. \ref{Fig_4}(b). 

In summary, we have used a DRL methodology to tackle the challenge of efficient navigation of colloidal robots in an unfamiliar complex environment. The proposed deep convolution neural network allows a colloidal robot agent to learn useful representations of visually rich high-dimensional sensory input and use them to derive generalizable navigation strategies. Although this algorithm is applied to a simplified self-propelled particle model, its model-free nature make it  adaptive to colloidal robots with different forms of dynamics \cite{Teeff_PRE_2008,LiuY_SoftMatter_2019} or more realistic physics (e.g., hydrodynamics \cite{Yang_JCP_2017,Bechin_RMP_2016}). The proposed DRL algorithm can also be extended to multi-agent system \cite{Tang_ACSNano_2016, Xie_SciRobot_2019, Charlesworth_PNAS_2019} to study complex emerging behaviors or to control multiple robots to cooperate on tasks and assemble to non-equilibrium machines and devices \cite{Yang_JHU_2017}.

\begin{acknowledgments}
Support from National Natural Science Foundation of China (Grant No. 11672161) is acknowledged.
\end{acknowledgments}

% \bibliography{basename of .bib file}

% ************** REFERENCES ************ %

% ************ END ************ %

\end{document}